\documentclass[prl,showpacs,final,twocolumn]{revtex4}
\usepackage{graphicx}
\begin{document}

\title{Atomically precise placement of single dopants in Si}
\author{S.~R.~Schofield}
\email{steven@phys.unsw.edu.au}
\author{N.~J.~Curson}
\author{M.~Y.~Simmons}
\author{F.~J.~Ruess}
\author{T.~Hallam}
\author{L.~Oberbeck}
\author{R.~G.~Clark}
\affiliation{Centre for Quantum Computer Technology, School of Physics, University of New South Wales, Sydney, NSW 2052, Australia}
\pacs{03.67.Lx, 68.37.Ef, 68.65.-k, 85.35.-p}

\begin{abstract}
We demonstrate the controlled incorporation of P dopant atoms in Si~(001) presenting a new path toward the creation of atomic-scale electronic devices. We present a detailed study of the interaction of PH$_3$ with Si~(001) and show that it is possible to thermally incorporate P atoms into Si~(001) below the H desorption temperature. Control over the precise spatial location at which P atoms are incorporated was achieved using STM H-lithography. We demonstrate the positioning of single P atoms in Si with $\sim$~1 nm accuracy and the creation of nanometer wide lines of incorporated P atoms.
\end{abstract}
\maketitle
 
The ability to control the location of individual dopant atoms within a semiconductor has enormous potential for the creation of atomic-scale electronic devices, including recent proposals for quantum cellular automata~\cite{sn-jap-99-4283}, single electron transistors~\cite{tu-ijcta-00-553} and solid-state quantum computers~\cite{ka-na-98-133}.   Current techniques for controlling the spatial extent of dopant atoms in Si rely on either ion implantation techniques, or dopant diffusion through optical or electron-beam patterned mask layers. While the resolution of these techniques continues to improve they have inherent resolution limits as we approach the atomic-scale~\cite{itrs-01}. The work presented here looks beyond conventional techniques to position P dopant atoms with atomic-precision by using scanning tunneling microscopy (STM) based lithography on H passivated Si~(001) surfaces~\cite{ly-apl-94-2010,sh-sc-95-1590} to control the adsorption and subsequent incorporation of single P dopant atoms into the Si~(001) surface. 

First, we show the controlled adsorption of PH$_3$ molecules to STM-patterned areas of H-terminated Si~(001) surfaces.  In these studies, we have used the H-terminated surface as a reference where the intrinsic surface periodicity can be observed to identify both adsorbed PH$_3$ molecules~\cite{wa-prb-94-4534} and the previously unobserved room temperature dissociation product, PH$_2$.  We then show, using low PH$_3$ dosed clean Si~(001) surfaces, that both of these room temperature adsorbates can be completely dissociated using a critical anneal, and more importantly, that this results in the substitutional incorporation of individual P atoms into the top layer of the substrate.  Finally, we combine these two results to demonstrate the spatially controlled incorporation of individual P dopant atoms into the Si~(001) surface with atomic-scale precision.  Of crucial importance to this final result is that the anneal temperature for P atom incorporation lies below the H-desorption temperature, so that the H-resist layer effectively blocks any surface diffusion of P atoms before their incorporation into the substrate surface.

\begin{figure}
\includegraphics{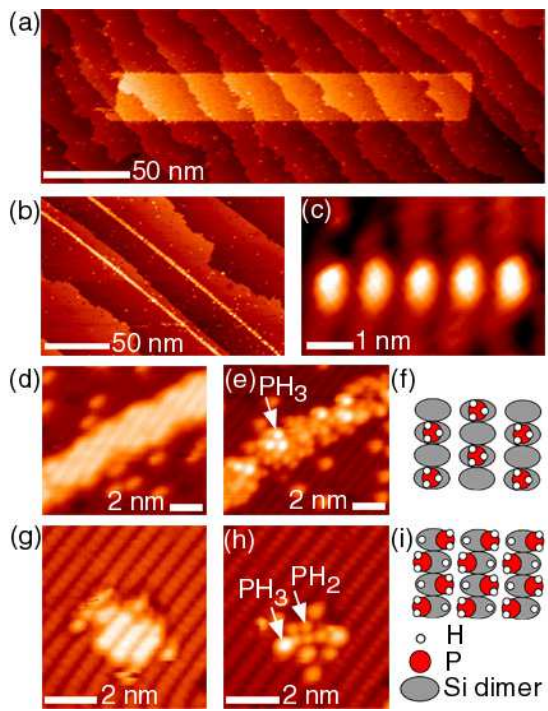}
\caption{A STM tip has been used to remove H atoms from a H-terminated Si~(001) surface to form (a), 200 $\times$ 30~nm$^2$ rectangular patch, (b), two parallel lines of exposed Si~(001) surface and (c), five single H-atom desorption sites. The desorption parameters used were +4~V sample bias and 1~nA tunnel current. The areas of bare Si~(001) surface appear brighter than the surrounding H-terminated surface due to the additional tunnel current contributed by the Si surface states~\cite{ha-prl-87-2071}. Images (d),(g) show a $\sim2$~nm wide lithographic line and $\sim2$~nm patch of hydrogen lithography.  After exposure to PH$_3$, the adsorbed PH$_3$ and PH$_2$ molecules can be seen within the lithographic areas in images (e) and (h) adsorbing with c(4 $\times$ 2) and p(2 $\times$ 2) periodicities, respectively, as shown in the schematics (f) and (i).}
\label{fig1}
\end{figure}

Figures~\ref{fig1}(a) - \ref{fig1}(c) demonstrate the flexibility of STM H-lithography to create different sized regions of bare Si~(001) surface.  As we will show, these regions can be used not only as a template for dopant incorporation but also to aid in fundamental studies of surface reactions.  Figures~\ref{fig1}(a) and \ref{fig1}(b) show the creation of both large areas (200 $\times$ 30~nm$^2$) and parallel, nanometer-wide lines of exposed Si, respectively.  The ultimate resolution limit is demonstrated in Fig.~\ref{fig1}(c) where we have desorbed single H atoms by pulsing the STM tip at $\sim$~0.75~nm intervals.  

In Figs.~\ref{fig1}(d) - \ref{fig1}(i) we confine the adsorption of phosphine to such lithographically defined areas of bare Si in order to identify the dissociation products PH$_3$ and PH$_2$.  Figure~\ref{fig1}(d) shows a $\sim$~2~nm wide line of exposed Si created using STM lithography on a H-terminated Si~(001) surface that was subsequently dosed with 0.3~Langmuir (L~$=0.75\times10^{-6}$~mbar~s) of PH$_3$ gas at room temperature (Fig.~\ref{fig1}(e)). Within the lithographic line in Fig.~\ref{fig1}(e) we can identify adsorbed PH$_3$ molecules by their appearance as circular protrusions centered on the dimer rows and ordered into a c(4$\times$2) periodicity (Fig.~\ref{fig1}(f))~\cite{wa-prb-94-4534}. However, PH$_3$ adsorption is known to be partially dissociative at room temperature~\cite{sh-jpc-96-4961,ha-cr-96-1261}, producing adsorbed PH$_2$ and H as well as PH$_3$.  Repeating this experiment at higher resolution using a smaller, $\sim$~2~nm diameter patch (Figs.~\ref{fig1}(g) and \ref{fig1}(h)), we have been able to identify these PH$_2$ molecules as adsorbate protrusions positioned asymmetrically about the dimer rows and ordered with a p(2$\times$2) periodicity, shown schematically in Fig.~\ref{fig1}(i).  In this situation we have actively used the intrinsic periodicity of the H-terminated surface to aid identification of the adsorbate species.  This identification is consistent with recent first principles calculations that predicted the dissociation fragments PH$_2$ and H should adsorb to the opposite side of the same dimer~\cite{mi-prb-01-125321}.  The above results demonstrate our ability to place and identify P precursor molecules at predefined locations on a Si surface. However, to create devices with full electrical activation we need to incorporate the P atoms from these molecules into the Si surface in substitutional lattice sites.  

\begin{figure}
\includegraphics{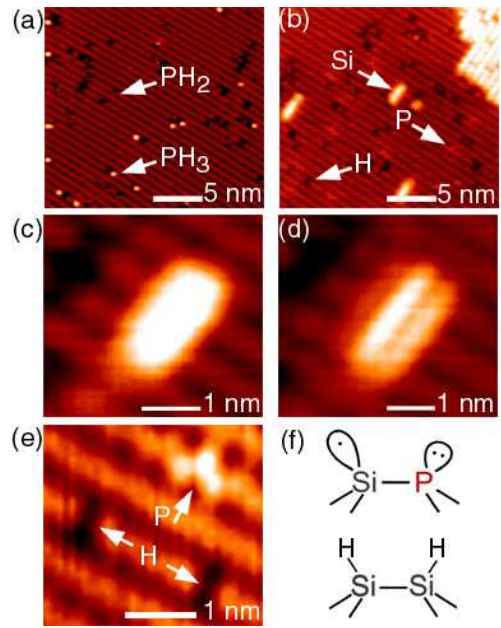}
\caption{STM images of a PH$_3$ dosed and annealed Si~(001) surface. (a), Si~(001) surface dosed with 0.01~L of PH$_3$ gas. The PH$_2$ molecules do not appear as bright as in Fig.~\ref{fig1}(h) due to the additional tunnel current produced by the Si surface $\pi$-states of the clean Si~(001) surface. (b), The PH$_3$ dosed surface after annealing to 350${}^\circ$C. (c), (d), Filled- and empty-state STM images of an ejected Si dimer chain. (e), A higher resolution image of the 350${}^\circ$C annealed surface showing a single P-Si heterodimer and two monohydride dimers. (f), Schematic diagrams of a P-Si heterodimer and a monohydride dimer. Parameters for all STM images were -1.6~V and 0.1~nA except (d), which was imaged at +1.6~V.}
\label{fig2}
\end{figure}

In previous studies, Hamers and coworkers~\cite{ha-ass-96-25} have suggested that individual P atoms are unstable on the Si~(001) surface and substitute for surface Si atoms to form P-Si heterodimers. For this process to occur, however, the adsorbed precursor PH$_3$ and PH$_2$ molecules must first completely dissociate into P and H.  For atomically precise placement of P in Si we require this incorporation process to take place without surface diffusion of the P atoms, or desorption of the adsorbed phosphine molecules or the surrounding H-resist layer.  To achieve this, we have first performed an extensive study of annealing lightly PH$_3$ dosed clean (H-free) Si~(001) surfaces.  Figure~\ref{fig2}(a) shows a STM image of a Si~(001) surface that has been flash-annealed to 1200${}^\circ$C, cooled to room temperature and dosed with 0.01~L PH$_3$. From the results shown in Figs.~\ref{fig1}(d) - \ref{fig1}(i), we are able to identify molecularly adsorbed PH$_3$ with an apparent height of $\sim$~0.07~nm above the substrate dimer rows (our data and Ref.~\onlinecite{wa-jpc-94-5966}), as well as PH$_2$ molecules by their distinctive asymmetric bonding arrangement.  Previous studies~\cite{ha-ass-96-25,ob-apl-02-3197} have observed the formation of P-Si heterodimers upon annealing PH$_3$ dosed surfaces to $\sim$~550${}^\circ$C.  However, at this temperature all H is desorbed from the surface making this anneal temperature incompatible with H-resist lithography.  Here, we use an anneal temperature of $\sim$~350${}^\circ$C, which we find is sufficient to incorporate P atoms, while being below the H-desorption temperature.

The most obvious feature of the 350${}^\circ$C annealed surface, Fig.~\ref{fig2}(b), is the appearance of bright, short, one-dimensional (1D) lines above the substrate. These 1D lines run perpendicular to the underlying dimer rows as would be expected for the next Si atomic layer (visible in the top right of Fig.~\ref{fig2}(b)). The apparent height ($\sim$~0.14~nm) and orientation of these 1D lines confirm that they are single dimer rows of epitaxial Si formed by the anisotropic agglomeration~\cite{mo-prl-89-2393} of Si atoms ejected from the substrate. This is further evidenced by the characteristic splitting of the ejected dimer chains between filled- and empty-state tunneling conditions~\cite{qi-prb-99-7293}, as shown in Figs.~\ref{fig2}(c) and \ref{fig2}(d). These ejected Si chains are not observable at higher anneal temperatures (above 400${}^\circ$C) since the Si atoms have enough thermal energy to migrate to step edges. 

We now focus on the P-Si heterodimers that are formed when P is incorporated into the Si. In Fig.~\ref{fig2}(b) these P-Si heterodimers are difficult to resolve since they are overshadowed by the very bright 1D Si chains on the upper atomic layer.  By examining a smaller area of the surface that is free from ejected Si, Fig.~\ref{fig2}(e), we are able to resolve the characteristic asymmetry of the P-Si heterodimer.  We measure a $\sim$~0.03~nm height increase of the P-Si heterodimer compared to the neighbouring Si dimers, which arises from the additional tunnel current due to the dangling bond of the Si atom of the P-Si heterodimer (Fig.~\ref{fig1}(f)). Additionally, the incorporated P atom induces static buckling of the neighbouring Si dimers, such that the heterodimer appears to extend further along the dimer row than a single dimer width. The characteristic asymmetric appearance of this feature provides a clear signature of a single P atom incorporated into the Si~(001) surface forming a single P-Si heterodimer. 

In addition, we also observe dark, single dimer vacancy-like features on the 350${}^\circ$C annealed surface (Fig.~\ref{fig2}(e)). These depressions are H-terminated Si dimers (monohydride, H-Si-Si-H), formed by the adsorption of H to the surface subsequent to the dissociation of the PH$_3$ and PH$_2$ molecules~\cite{ha-cr-96-1261,bo-prl-91-1539}. Their dark appearance in STM images is characteristic of H passivating the surface and removing the Si $\pi$-states from the band gap~\cite{ha-cr-96-1261}. 

Identification of each of the various features on the surface in Fig.~\ref{fig2}(b) is further confirmed by STM studies of the surface at higher temperatures. Raising the anneal temperature (to $\sim$~450${}^\circ$C) causes the disappearance of the short 1D Si dimer chains from the surface as a result of the diffusion of the Si atoms from the ejected Si dimer chains to step edges, as observed in Si growth experiments~\cite{vo-ssr-01-127,dab-ch3}. Further annealing to $\sim$~550${}^\circ$C results in the disappearance of the monohydride dimers, as the monohydride is desorbed from the Si~(001) surface~\cite{su-ass-94-449}. Finally, annealing to $\sim$~750${}^\circ$C causes the removal of the P-Si heterodimers in agreement with the established desorption temperature of P from the Si (001) surface~\cite{co-jvsta-94-2995}.  These studies demonstrate that P atoms can be incorporated into the Si~(001) surface from adsorbed PH$_3$ molecules by annealing to 350${}^\circ$C and that this process is accompanied by the displacement of Si atoms onto the substrate and the adsorption of H. We have also confirmed the characteristic asymmetric appearance of isolated P atoms in this surface. 

\begin{figure}
\includegraphics{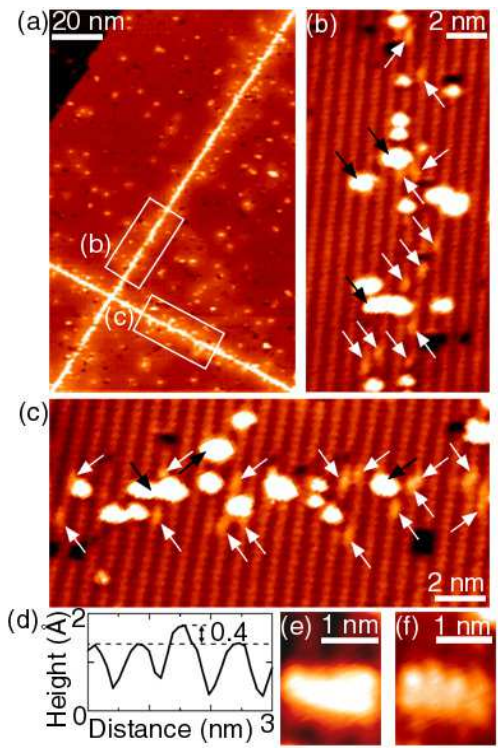}
\caption{STM images of controlled P incorporation into Si~(001) along two perpendicular $\sim$~4~nm wide lithographic lines. (a), Lithographic lines created in a H-terminated Si~(001) surface by removing H from the surface with an STM tip. These desorption lines were created by moving the tip with a velocity of 40~nm/s while applying tunneling conditions of +4~V sample bias and 1~nA tunnel current; (b), (c), High-resolution STM images of the boxed regions in image (a) after dosing with 0.3~L of PH$_3$ and annealing to 350${}^\circ$C.  Incorporated P atoms in the form of P-Si-H ``hydrided heterodimers'' are labelled by white arrows, while ejected Si dimer chains are labelled with black arrows; (d), Line profile taken over a single P-Si-H heterodimer, showing a $\sim$~0.04~nm height increase due to tunneling from the P lone pair orbital; (e), (f), Filled- and empty-state STM images of a H-terminated ejected Si dimer chain.}
\label{fig3}
\end{figure}

We now wish to control the spatial location of the P atom incorporation using the H-resist lithography technique described earlier for the creation of atomic-scale devices. Figure~\ref{fig3}(a) shows an STM image of a H-terminated Si~(001) surface where the STM tip has been used to selectively remove H atoms from Si along two perpendicular $\sim$~4~nm wide lines.  The patterned surface was then exposed to $\sim$~0.3~L PH$_3$ and annealed to $\sim$~350${}^\circ$C as shown in Figs.~\ref{fig3}(b) and \ref{fig3}(c) (which correspond to the areas of the surface marked by boxes in Fig.~\ref{fig3}(a)). Along the length of both lithographic lines, many asymmetric features are observed, corresponding to the formation of P-Si heterodimers (indicated by white arrows in Figs.~\ref{fig3}(b) and \ref{fig3}(c)). In contrast to the clean surface results, these P-Si heterodimers are H-terminated (P-Si-H) resulting from the almost complete re-termination of the lithographic line with H during the 350${}^\circ$C anneal.  Fig.~\ref{fig3}(d) shows a line profile taken over a single P-Si-H heterodimer, showing a $\sim$~0.04~nm increase due to tunneling from the lone pair state of the P atom. Importantly, we find that these P-Si-H heterodimers are solely located within the $\sim$~4~nm wide region of the original lithographic line, confirming that the H-resist layer, which survives the 350${}^\circ$C anneal intact, acts as an effective mask for the controlled incorporation of P into the surface.  We also observe clear evidence of ejected Si dimer chains along (and confined to) the lithographic line, as indicated by the black arrows in Figs.~\ref{fig3}(b) and \ref{fig3}(c). As was the case in the clean surface experiments, filled- and empty-state images of the ejected Si chains (Fig.~\ref{fig3}(e), \ref{fig3}(f)) reveal the characteristic splitting of these ejected Si dimer chains under empty-state imaging.  These results demonstrate that it is possible, using an appropriate anneal, to controllably incorporate P atoms from adsorbed phosphine molecules into Si~(001) at precise locations using STM-based H-lithography.

We note that the PH$_3$ dose rate and fluence is critical in obtaining the successful incorporation of P within our lithographic H-desorption sites and not at random single dimer H desorption sites (which are sometimes present due to incomplete H passivation during the H-dosing stage).  Using both higher PH$_3$ dose rates (10$^{-8}$ mbar chamber pressure) and total doses ($\sim$~3 L), we have noted the adsorption of PH$_3$ at single H desorption sites~\cite{ob-prb-01-161401}. However, with the low dose rate and fluence used here (10$^{-9}$ mbar and 0.3 L) we have observed an almost complete absence of adsorption at single H desorption sites, with coverages approaching saturation in the larger lithographically defined desorption regions. 

\begin{figure}
\includegraphics{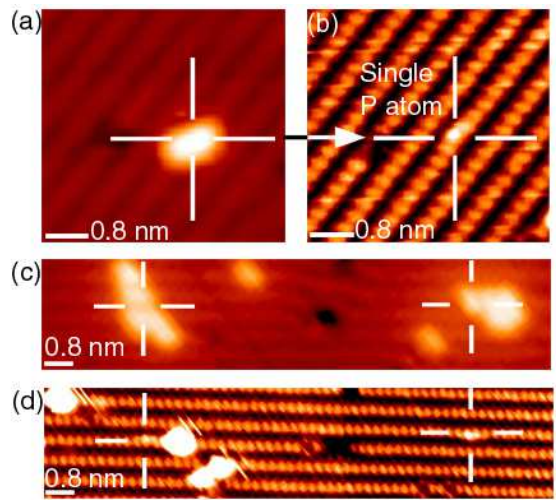}
\caption{STM images of atomically-controlled single P atom incorporation into Si~(001). (a), H-terminated Si~(001) with a $\sim$~1~nm diameter H-desorption point ($\sim$3 dimers long). (b), The same area after PH$_3$ dosing and annealing to 350${}^\circ$C showing a single P atom incorporated at the location defined by the H-desorption point. (c), Two H-desorption sites separated by $\sim$~12~nm. (d), The same area of the surface shown in (c), after PH$_3$ dosing and annealing to 350 ${}^\circ$C.  A single P atom has been incorporated into the surface within both of the two lithographically defined areas.  The separation of these two P atoms ($\sim$~12~nm) is of the order of the 20~nm separation required for the P qubit array of the Kane solid-state quantum computer~\cite{ka-na-98-133}.}
\label{fig4}
\end{figure}

Finally, we have been able to achieve the ultimate limit of P atom placement in Si~(001) by incorporating single P atoms into the Si~(001) surface with atomic-scale precision. Figure~\ref{fig4}(a) shows a H-terminated Si surface with a controlled desorption site of diameter $\sim$~1~nm, exposing just two or three of the substrate Si dimers. Figure~\ref{fig4}(b) shows the same surface area after dosing with $\sim$~0.3~L PH$_3$ and annealing to $\sim$~350${}^\circ$C. Here we see a very clear signature of the asymmetric P-Si-H heterodimer within the $\sim$~1~nm lithographic area, and complete re-termination of this area with H. This is the first demonstration of the spatially controlled incorporation of a single dopant atom in Si and can be produced repeatably in our laboratory.  In Figs.~\ref{fig4}(c) and \ref{fig4}(d) we have incorporated two P atoms in Si $\sim$~12~nm apart for the qubit array of a Si-based quantum computer~\cite{ka-na-98-133}.  Figure~\ref{fig4}(c) shows two desorption sites of diameter $\sim$1~nm and separated by $\sim$12~nm.  After PH$_3$ dosing and annealing we observe the incorporation of a single P atom at each of these lithographic sites, as seen in Fig.~ \ref{fig4}(d).  We note the presence of two single H atom desorption sites in addition to the two lithographically defined desorption regions in Fig.~\ref{fig4}(c).  However, as seen in Fig.~\ref{fig4}(d), these sites do not produce incorporated P atoms at the low PH$_3$ dose rate and fluence used.  Additionally, in Fig.~\ref{fig4}(d), several single H atom desorption sites are seen as white protrusions surrounding the left P-Si-H heterodimer.  These are the result of single H atom desorption events during the 350${}^\circ$C anneal and do not affect the P atom placement.  While the accuracy required for qubit placement in a solid-state quantum computer depends on the final separation required for electron wavefunction overlap, the $\sim$~1~nm placement accuracy that we demonstrate here is more than adequate for the Kane qubit architecture~\cite{ka-na-98-133}. 
In summary, we have demonstrated the first controlled incorporation of single P dopant atoms in Si with atomic-scale precision. To elucidate this, we have first studied the interaction of PH$_3$ with the Si~(001) surface and in particular observed the substitutional incorporation of P atoms into the substrate after annealing to 350${}^\circ$C.  We then employed STM-based H-lithography to control the spatial locations of P atoms incorporated into the substrate.  Critical to this result is that the 350${}^\circ$C anneal lies well below the H-desorption temperature such that the H-resist layer acts as an atomic scale mask for controlled P atom incorporation.  Preliminary studies to encapsulate these dopants in Si---and observe the resulting change in surface electrostatic potential~\cite{mo-apl-03-1932} to confirm minimal segregation of array---are underway with promising results~\cite{ob-srl-02-00}. These results open the door to the exciting possibility of creating electronic devices in Si with atomically controlled dopant profiles, such as atomically ordered transistors, quantum cellular automata, single atom memory devices and the fabrication of a solid-state quantum computer.

\end{document}